# Impact of Strain on Drain Current and Threshold Voltage of Nanoscale Double Gate Tunnel Field Effect Transistor: Theoretical Investigation and Analysis

Sneh Saurabh and M. Jagadesh Kumar*

Department of Electrical Engineering, Indian Institute of Technology, Delhi, Hauz Khas, New Delhi 110 016, India

Tunnel field effect transistor (TFET) devices are attractive as they show good scalability and have very low leakage current. However they suffer from low on-current and high threshold voltage. In order to employ the TFET for circuit applications, these problems need to be tackled. In this paper, a novel lateral strained double-gate TFET (SDGTFET) is presented. Using device simulation, we show that the SDGTFET has a higher on-current, low leakage, low threshold voltage, excellent subthreshold slope, and good short channel effects and also meets important ITRS guidelines.

_______________________________________________

* mail address: mamidala@ieee.org

# 1. Introduction

As the semiconductor devices are miniaturized, high leakage currents and short channel effects become significant enough to be the major concerns for circuit designers. Many novel device designs have been suggested in order to tackle this problem. Different types of transistors that utilize the band-to-band tunneling as the basic operating principle have been demonstrated in the literature[1-17]. These transistors offer advantages in terms of very low leakage current, good sub-threshold swing, improved short channel characteristics and lesser temperature sensitivity. However the greatest challenge in adopting tunnel devices for wide-scale application is its low on-current. A high threshold voltage is another practical problem that needs to be tackled. Some of the techniques that address these issues are: vertical tunnel field effect transistor (TFET) with SiGe delta doped layer[14], tunnel bandgap modulation, gate work function engineering[15], high-k gate dielectric with double gate[17], higher source doping and abrupt doping profile[10]. In order to make TFET suitable for wide-scale application, the device parameters have to meet the ITRS guidelines[18]. Also the device structure should be such that it can be incorporated in the existing process flow with not much modification.

In this paper, we propose a novel lateral strained double gate TFET (SDGTFET). Using two-dimensional simulation, we demonstrate that the proposed structure exhibits a pragmatic threshold voltage, improved on-current, very good subthreshold slope, high on-current to off-current ratio and negligible short channel effects and meets the ITRS near-term guidelines. We have provided the physical reasoning for this improved performance and also given guidelines for device optimization. Since the device structure presented here is similar to the conventional strained double gate FET (DGFET) technology, it can be easily integrated into the existing fabrication technology.

## 2. Device Structure

The cross-sectional view of SDGTFET is shown in Fig. 1. This structure is similar to the conventional double gate TFET (DGTFET) except that the complete silicon body is strained silicon which can be fabricated using single-layer strained-silicon-on-insulator (SSOI) technology[19-21]. SSOI is a novel SiGe-free material system that has the advantages of strained silicon while improving the scalability of thin-film SOI. The amount of strain in an SSOI is controlled by varying the mole fraction of Ge in the relaxed SiGe buffer layer that is used during its fabrication.

The source and drain in a TFET are heavily doped as it facilitates greater tunneling. The source is doped $P^+$ with $1x10^{20}$ atoms/cm$^3$ and drain is doped $N^+$ with $5x10^{18}$ atoms/cm$^3$. The channel region is doped N-type with $1x10^{17}$ atoms/cm$^3$. In the normal mode of operation of this device, the source is grounded and a positive voltage is applied to the drain. The gate voltage controls the tunneling by modulating the carrier concentration in the channel region. A lower doping is kept on the drain side so that the tunneling is suppressed at the drain when a negative voltage is applied to the gate[10]. This is required to realize an NMOS type of operation. However, if a PMOS type of operation is required doping can be made higher on the drain side. In this paper, we mainly concentrate on NMOS type of operation. The device parameters used in our simulation are given in Table 1. The gate work function is taken as 4.5 eV corresponding to a metal gate stack[17].

## 3. Simulation Model and Operating Principle

All simulations have been carried out using Silvaco's device simulator ATLAS version 1.12.1.R[22]. The Hurkx band-to-band tunneling model has been used in this

work[23,24)]. Since the tunneling process is non-local and it is necessary to take into account the spatial profile of the energy bands, non-local band-to-band tunneling model was used. As the source and drain are heavily doped, the band gap narrowing effect is also taken into account. The drift-diffusion model of current transport is used in the simulations. In order to validate our device simulations and the choice of tunneling model, we first simulated the TFET structure[17] and calculated the band-structure and the transfer characteristics. Our simulation results matched with those shown in Figs. 2 and 4(a) reported in ref. 17.

In Fig. 2, the energy band diagram for the SDGTFET is shown for two Ge mole fraction values. Fig. 2(a) shows the band diagram of an SDGTFET for zero Ge mole fraction implying an unstrained silicon DGTFET. It can be seen that when the device is in the off state ($V_{GS} = 0$ V), there exists a large barrier at the source end of the device which inhibits the tunneling phenomenon and hence the current flow between the source and the drain. However when the gate voltage is increased to 1 V, the energy bands at the source end of the device get aligned and the tunneling barrier width decreases drastically. This increases the tunneling probability of the electrons from the valence band in the source to the conduction band in the channel and forms the basis of operation for a tunneling transistor. Figure 2(b) shows the band diagram of an SDGTFET when the Ge mole fraction is increased to 0.5. When the gate voltage is increased to 1 V, a similar kind of band alignment takes place in an SDGTFET (x = 0.5) which facilitates the tunneling phenomenon. However, the tunneling width is reduced considerably more in an SDGTFET (x =0.5) compared to a conventional DGTFET. In order to make this point clear, the tunneling width of an SDGTFET was extracted from the energy band diagram and shown in Fig. 2(c). The horizontal distance of the top of the valence band to the conduction band as shown in Fig. 2(a) is taken as the tunnel width. It is clear from Fig.

2(c) that the tunneling width gets reduced with the increase in Ge mole fraction. Since the tunneling probability increases due to the reduced tunneling width, the current of an SDGTFET is expected to increase with increasing Ge mole fraction. The dependence of tunneling current on various device and material parameters can be described by[17, 25]

$$I \propto \exp\left(-\frac{4\sqrt{2m^*}E_g^{3/2}}{3|e|h(\Delta\Phi + E_g)}\sqrt{\frac{\varepsilon_{Si}}{\varepsilon_{ox}}t_{ox}t_{Si}}\right)\Delta\Phi,$$ (1)

where $m^*$ is the effective carrier mass, $E_g$ is the band gap, $\Delta\Phi$ is the energy range over which tunneling can take place, $t_{ox}$, $t_{Si}$, $\varepsilon_{ox}$, and $\varepsilon_{Si}$ are the oxide and silicon film thicknesses and dielectric constants, respectively, $e$ is the electronic charge and $h$ is the Planck's constant[17]. Equation (1) shows that the tunneling current can be modulated by changing the band-gap of silicon. This inference is the primary motivation to introduce strained-silicon in an SDGTFET. The presence of strain causes the bandgap and the effective mass of carriers in silicon to decrease and the electron affinity of silicon to increase. This can be modeled as[26-28]

$$\left(\Delta E_g\right)_{s-Si} = 0.4x,$$ (2)

$$\left(\Delta E_C\right)_{s-Si} = 0.57x,$$ (3)

$$V_T \ln\left(\frac{N_{V,Si}}{N_{V,s-Si}}\right) = V_T \ln\left(\frac{m^*_{h,Si}}{m^*_{h,s-Si}}\right)^{\frac{3}{2}} \approx 0.075x,$$ (4)

where $x$ is the Ge mole fraction in the relaxed SiGe buffer layer; $\left(\Delta E_g\right)_{s-Si}$ is the decrease in the bandgap of silicon due to strain; $\left(\Delta E_C\right)_{s-Si}$ is the increase in electron affinity of silicon due to strain; $V_T$ is the thermal voltage; $N_{V,Si}$ and $N_{V,s-Si}$ are the density of states (DOS) in the valence band in the normal and strained silicon, respectively; $m^*_{h,Si}$ and $m^*_{h,s-Si}$ are the hole DOS effective masses in normal and strained-silicon, respectively. Using eqs. (2) and (3), for a given $x$, the simulator calculates the change in bandgap and the electron affinity for the strained silicon. Using the standard values of $N_{V,Si}$ and $m^*_{h,Si}$ given in ref. 21 and using eq. (4), the simulator calculates the change in the effective density of states and the hole DOS effective mass in the strained silicon for a given x. The effect of warping and nonparabolicity are not considered in our simulations. Also, we have taken the tunnel effective mass in strained silicon same as in normal unstrained silicon.

In the next section, using two-dimensional simulation, we show how the modulation of band-structure of silicon (engineered by straining) results in an overall improvement of the device characteristics of an SDGTFET and hence makes it capable to meet the ITRS near-term guidelines.

## 4. Simulation Results

### 4.1 Transfer characteristics of DGTFET

To highlight the advantages and short-comings of a DGTFET, we have first simulated and compared the transfer characteristics of a DGTFET with a conventional DGFET as shown in Fig. 3. The ITRS near-term guideline for low standby-power devices is also shown in this figure. Firstly, we observe that the off-current of a DGTFET is very low compared to a DGFET and is also significantly below the ITRS guideline. This is the biggest advantage of the tunneling devices as compared to the DGFET whose off-state current is close to violating the ITRS guideline for low standby power devices. However, the on-current of a DGTFET is drastically low compared to a DGFET and is also significantly below the ITRS guideline. Therefore, the DGTFET cannot be used as a replacement for the conventional devices unless the on-current of the DGTFET is increased by a few orders of magnitude. SDGTFET proposed in this work provides a viable solution to achieve this without significantly affecting the off-state leakage current. Another advantage of a DGTFET over conventional DGFET can be noticed in Fig. 3. The subthreshold slope of the DGFET of Fig. 3 is around 70 mV/decade (for an ideal DGFET it is 60 mV/decade). In a DGTFET, the subthreshold slope is a strong function of the gate voltage. At low gate voltage, when the drain current starts increasing (0.3 V), the subthreshold slope of a DGTFET is around 33 mV/decade, which is much better than an ideal MOSFET. This indicates a very steep rise in drain current with the gate voltage (at low gate voltages). However, as the gate voltage increases, the subthreshold slope degrades, but it still remains better than that of a DGFET (70 mV/decade) up to around 0.6 V. This steep rise in drain current with gate voltage is a one of the major advantages of a DGTFET over conventional DGFET.

*4.2 Transfer characteristics of SDGTFET*

In order to study the effect of strain on the behavior of the DGTFET, transfer characteristics of the SDGTFET at different Ge mole fractions were computed and shown in Fig. 4. The ITRS guideline for the off-current and the on-current are also shown in the same figure. It can be seen that the on-current of the device improves appreciably as the Ge mole fraction is increased. The on-current increases by around 2 orders of magnitude for a Ge mole fraction of 0.5 and also meets the ITRS requirements. It should be noted that while the on-current increases with strain, the off-current (defined as the drain current when $V_{GS} = 0$ V and $V_{DS} = 1$ V) also increases. However the off-current is still much below the ITRS near-term guideline for low-standby power technology as can be seen in Fig. 4. The ratio of on-current and off-current at various Ge mole fractions is shown in Fig. 5(a). This shows that the ratio of on-current to off-current goes through a maximum value with an increase in Ge mole fraction. The primary reason for it is that the evolution of on-current and off-current are not similar with the increase in Ge mole fraction. The total current in a tunnel FET is composed of two components: the band-to-band tunneling current and the reverse biased diode current[14]. At low gate voltages (when the TFET is in weak inversion region), the tunneling probability is low and the band-to-band tunneling and the reverse biased diode current both decide the off-current. But at high gate voltages (when the TFET is in strong inversion) the band-to-band tunneling current is dominant and solely governs the on-current of the device[14]. Also, the evolution of the tunneling current of an SDGTFET in off-sate with the increase in Ge mole fraction is not same as the evolution of tunneling-current in on-state. Figure 5(b) shows the band diagram of an SDGTFET in off state ($V_{GS}$ = 0 V) at different Ge mole fractions (x=0.0 and 0.5). It can be seen that at a lower Ge mole fraction (x = 0.0), the tunneling barrier width at the source side is quite large compared with the tunnel width on the drain side. Therefore, the tunneling current on the source side is almost negligible at lower mole fractions. However, as the Ge mole fraction is increased (x = 0.5), the tunneling width on the source side decreases drastically while the tunneling width on the drain side remains almost constant. This results in an almost no increase in off-current at lower Ge mole fractions (till the tunneling at the source side becomes significant) and greater increase in off-current at higher Ge mole fractions (when the tunneling on the source side becomes appreciable). This shows that the strain in the device can be engineered to get the desired ratio of on-current and off-current.

It should be noted that strain also increases the mobility of the carriers[29]. But in a TFET, mobility may not play a major role. This was verified by varying the mobility in the simulation model and the transfer characteristics were found to be almost independent of the mobility as shown in Fig. 6. The drift-diffusion mechanism of current transfer, which is the dominant mechanism of current transfer in a normal DGFET, is not important in the case of a TFET. The increase in tunneling current with strain in the TFET can therefore be attributed to the decrease in barrier width with strain as shown in Fig. 2(b). However it may be noted that the effect of mobility is less pre-dominant only when the device on-current, limited by tunnel injection, remains much lower than that of a standard MOSFET. In fact, Fig. 6 shows that the turn on-current of the DGTFET is roughly 5-10 % the on-current of the standard DGFET at the same dimensional size. If the current is not limited by tunnel injection, then it would be limited by carrier mobility, as the two effects happen to be in series.

It should also be noted that strain affects tunneling current through various physical parameters as shown in eqs. (1)-(4). However the increase in drain current in an SDGTFET is mostly brought about by the changes in the band-structure in strained silicon. This was verified by artificially changing only the band-structure in normal unstrained silicon and keeping all other physical parameters unchanged in the simulation. It was found that there was still around 90% improvement in the drain current just due to change in the band-structure. Therefore, it may be concluded that the improvement in on-current in an SDGTFET is mostly due to the change in the band-structure in the strained silicon.

Another important observation is that the SDGTFET can derive the advantages of strain both in NMOS and PMOS type of operation. In this paper, we have discussed only NMOS mode of operation. In an NMOS type of TFET, when a positive gate voltage is applied to the gate, electrons tunnel from the valence band ($p^+$ doped region) to the conduction band in the channel and then flow to the $n^+$ doped region. Since $p^+$ doped region is the source of electrons, this is named as the source terminal and the $n^+$ region is named as drain[10]. When the gate voltage is made negative, tunneling can take place on the drain side as well. In an NMOS type of TFET this tunneling is suppressed by keeping a lower doping on the drain side (n$^+$ region). However, in order to make this device predominantly PMOS type, the doping of the n$^+$ region is increased and the doping of the p$^+$ region is made lower. When a negative gate voltage is applied, electrons can tunnel from the valence band (channel region) to the conduction band in the n$^+$ region. The generated holes flow to the p$^+$ doped region. A higher doping at the n$^+$ region ensures an enhanced tunneling when a negative gate voltage is applied. In a PMOS type of TFET, the n$^+$ region is referred to as source and the p$^+$ region is referred to as drain[10]. Since strain exists throughout the silicon body, the current would be greater in PMOS operation also. Figure 7 shows the transfer characteristics of an SDGTFET when operating in both of these modes. As it is evident from the figure, the transfer characteristics of this device are almost symmetrical and hence the W/L scaling for PMOS and NMOS need not be done for this device. This is one of the advantages of using totally strained silicon instead of delta-doped SiGe on the source side as in[14, 15].

*4.3 Threshold voltage*

We have used the constant current method to define the threshold voltage[14]. The gate voltage at which the drain current becomes $1 \times 10^{-7}$ A/μm is taken as the threshold voltage. For an SDGTFET with no strain (Ge mole fraction of 0), the threshold voltage is around 0.9 V as it can be inferred from Fig. 3. This is much higher than the ITRS guideline. ITRS near term guideline for low standby power technology sets threshold voltage close to 0.3 V[25]. When strained silicon is used, the threshold voltage is reduced. This is because of increased tunneling at a given gate voltage due to a reduced tunneling width as shown in Fig. 2(b). Figure 8 shows the threshold voltage of the device at various strains corresponding to different Ge mole fraction for different work functions. As it can be seen, for higher strain the threshold voltage is reduced and meets the ITRS guidelines. However, as expected decreasing the gate work-function also increases the off-current and hence reduces the ratio of on-current to off-current. Figure 9(a) shows how the ratio of on-current to off-current changes with change in Ge mole fraction at different work-functions. For a small work function (e.g., 4.2 eV), the threshold voltage will be small and it will further decrease with increasing Ge mole fraction. This will result in an increase in off-state current $I_{off}$ calculated at $V_{GS}$=0 leading to a decreased ratio of on-current to off-current. Therefore, an intelligent trade-off needs to be made between the Ge mole fraction and the work function of the device in order to get a desired threshold voltage while maintaining an acceptable ratio of on-current to off-current. Figure 9(b) illustrates how such a tradeoff can be made. Here, it is assumed that the ratio of on-current to off-current has to be kept greater than $1 \times 10^8$. Figure 9(b) shows the combination of the Ge mole fraction and the work-function of the gate that can be used to get the required threshold voltage while satisfying the above mentioned constraint on the ratio of off-current to on-current. These curves are plotted by computing the threshold voltage at various Ge mole fractions and gate work functions and then imposing the constraint on the ratio of on-current to off-current using Fig. 9(a). It is worthy to note that an SDGTFET at higher Ge mole fraction (greater than x = 0.4) is capable of meeting the ITRS near term low-power guideline for threshold voltage (around 0.3 V) while satisfying a reasonable constraint on the ratio of on-current to off-current.

*4.4 Subthreshold swing*

The subthreshold swing of a TFET is not limited to 60 mV/decade as for the normal MOSFETs. It has been theoretically and experimentally proven that a lower subthreshold swing for a TFET can be realized[9,10,30]. This is one of the major advantages of a TFET over conventional devices. Since the subthreshold slope of a TFET is a strong function of gate voltage, the average subthreshold swing of the device has to be considered. Average subthreshold swing is defined as[17]

$$S = \frac{V_t - V_{off}}{\log I_{vt} - \log I_{off}}, \qquad\qquad\qquad (5)$$

where $V_t$ is the threshold voltage, $V_{off}$ is the voltage at which device is off, $I_{vt}$ is the drain current at threshold and $I_{off}$ is the off current of the device. The off-current $I_{off}$ is defined as the drain current when gate voltage is zero[17]. To make sure that computational noises do not affect the accuracy of the calculated $I_{off}$, a very fine mesh is used in our simulations at the junctions and more particularly across the region where tunneling takes place. Average subthreshold swing measures the amount of gate voltage required (on an average) to increase the device current by a decade in the subthreshold region. Average subthreshold swing is a crucial parameter that affects the performance of the device as a switch[17]. Therefore we have considered this parameter for benchmarking an SDGTFET.

When the normal silicon is replaced by strained silicon, the subthreshold swing improves further. Figure 10 shows the average subthreshold swing of the device at various Ge mole fractions. At a Ge mole fraction greater than 0.4, the subthreshold swing of an SDGTFET is even better than the ideal normal MOSFET.

*4.5 Threshold Voltage Roll-Off*

As the channel length is decreased, threshold voltage roll-off is one of the significant problems. It is found that the threshold voltage roll-off for an SDGTFET is almost negligible. Figure 11 shows the threshold voltage roll off of an SDGTFET at different mole fractions of Ge. As it can be seen from the figure, there is no appreciable decrease in threshold voltage up to 20 nm, both for strained and unstrained DGTFET. Though it may be expected that the strained silicon may show worse threshold voltage roll-off (as it has a lower band-gap), no such effect was observed. The tunneling phenomenon in both a DGTFET and in an SDGTFET is confined to a very small region around the source. Therefore reducing the gate length does not show any major impact on the performance of this device until the drain is brought too close to source so as to impact the tunneling process. It should be noted that a double gate structure and thin SOI channel are known to show a better scalability. However the improved scalability in an SDGTFET cannot be solely attributed to this. Figure 11 also shows the threshold voltage roll-off for a similar DGFET (silicon film thickness=10nm, gate oxide thickness = 3nm and $\Phi_m$= 4.71 eV) computed using a constant current method. It can be seen that in the case of a DGFET, the threshold voltage begins to fall at a channel length of around 70 nm, while for an SDGTFET this fall begins at around 20 nm. Therefore it shows that the scalability advantage for an SDGTFET is not just because of double gate and thin body, but due to the tunneling phenomenon. Therefore, an SDGTFET is very immune to threshold voltage roll-off and hence can be a very good candidate as the channel length of the transistors is reduced.

## 5. Conclusion

In this paper, we have presented the effect of strain on the performance of double gate TFET device structure. We have demonstrated that using this strained DGTFET, it is possible to meet important ITRS guidelines such as the on-current and threshold voltage. We have also shown that although the off-current increases with strain, it is still much below the ITRS guideline. Further, the device can be engineered using strain to get a particular ratio of on-current and off-current. The subthreshold swing is excellent and short channel effects of this device are low. Also the structure is compatible with strained DGFET fabrication technology and can be seamlessly integrated into it. The device parameters can be optimized so as to get the best on-current, on-current to off-current ratio, threshold voltage, subthreshold slope and short-channel effects. However, it may be pointed out that the problem of high on-resistance at a low drain-source voltage still needs to be tackled in tunneling FETs. As the device size gets reduced further and leakage requirements become more stringent, this device can certainly be one of the best alternatives[31].

REFERENCES

1) P. F. Wang, T. Nirschl, D. S. Landsiedel, and W. Hansch: Solid-State Electron. **47** (2003) 1187.

2) W. Hansch, C. Fink, J. Schulze, and I. Eisele: Thin Solid Films **369** (2000) 387.

3) S. Sedlmaier, K. K. Bhuwalka, A. Ludsteck, M. Schmidt, J. Schulze, W. Hansch, and I. Eisele: Appl. Phys. Lett. **85** (2004) 1707.

4) V. Dobrovolsky, V. Rossokhaty, and S. Cristoloveanu: Solid-State Electron. **50** (2006) 754.

5) W. M. Reddick and G. A. J. Amaratunga: Appl. Phys. Lett. **67** (1995) 494.

6) T. Uemura and T. Baba: Solid-State Electron. **40** (1996) 519.

7) C. Aydin, A. Zaslavsky, S. Luryi, S. Cristoloveanu, D. Mariolle, D. Fraboulet, and S. Deleonibus: Appl. Phys. Lett. **84** (2004) 1780.

8) T. Nirschl, S. Henzler, J. Fischer, M. Fulde, A. B. Stoffi, M. Sterkel, J. Sedlmeir, C. Weber, R. Heinrich, U. Schaper, J. Einfeld, R. Neubert, U. Feldmann, K. Stahrenberg, E. Rudrer, G. Georgakos, A. Huber, R. Kakoschke, W. Hansch, and D. S. Landsiedel: Solid-State Electron. **50** (2006) 44.

9) Q. Zhang, W. Zhao, and A. Seabaugh: IEEE Electron Device Lett. **27** (2006) 297.

10) P. F. Wang, K. Hilsenbeck, T. Nirschl, M. Oswald, C. Stepper, M. Weis, D. S. Landsiedel, and W. Hansch: Solid-State Electron. **48** (2004) 2281.

11) S. Tehrani, J. Shen, H. Goronkin, G. Kramer, R. Tsui, and T. X. Zhu: IEEE Electron Device Lett. **16** (1995) 557.

12) J. Shen, S. Tehrani, H. Goronkin, G. Kramer, and R. Tsui: IEEE Electron Device Lett. **17** (1996) 94.

13) K. K. Bhuwalka, S. Sedlmaier, A. K. Ludsteck, C. Tdorf, J. Schulze, and I. Eisle: IEEE Trans. Electron Devices **51** (2004) 279.

14) K. K. Bhuwalka, J. Schulze, and I. Eisele: IEEE Trans. Electron Devices **52** (2005) 1541.

15) K. K. Bhuwalka, J. Schulze, and I. Eisele: IEEE Trans. Electron Devices **52** (2005) 909.

16) V. V. Nagavarapu, R. R. Jhaveri, and J. C. S. Woo: IEEE Trans. Electron Devices **55** (2008) 1013.

17) K. Boucart and A. M. Ionescu: IEEE Trans. Electron Devices **54** (2007) 1725.

18) Semiconductor Industry Association (SIA), International Technology Roadmap For Semiconductors, 2006 Update.

19) H. Yin, K. D. Hobart, R. L. Peterson, F. J. Kub, S. R. Shieh, T. S. Duffy, and J. C. Sturm: IEDM Tech. Dig., 2003, p. 3.2.1.

20) K. Rim, K. Chan, L. Shi, D. Boyd, J. Ott, N. Klymko, F. Cardone, L. Tai, S. Koester, M. Cobb, D. Canaperi, B. To, E. Duch, I. Babich, R. Carruthers, P. Saunders, G. Walker, Y. Zhang, M. Steen, and M. Ieong: IEDM Tech. Dig., 2003, p. 3.1.1.

21) S. Takagi, T. Mizuno, T. Tezuka, N. Sugiyama, T. Numata, K. Usuda, Y. Moriyama, S. Nakaharai, J. Koga, A. Tanabe, and T. Maeda: Appl. Surf. Sci. **224** (2004) 241.

22) ATLAS User's Manual, Silvaco International, Santa Clara, CA, 2008.

23) G. A. M. Hurkx, D. B. M. Klassen, and M. P. G. Knuvers: IEEE Trans. Electron Devices **39** (1992) 331.

24) G. A. M. Hurkx, D. B. M. Klassen, M. P. G. Knuvers, and F. G. O. Hara: IEDM Tech. Dig., 1989, p. 307.

25) J. Koch and J. Appenzeller: Proc. 63rd DRC, 2005, p. 153.

26) T. Numata, T. Mizuno, T. Tezuka, J. Koga, and S. Takagi: IEEE Trans. Electron Devices **52** (2005) 1780.

27) J. S. Lim, S. E. Thompson, and J. G. Fossum: IEEE Electron Device Lett. **25** (2004) 731.

28) W. Zhang and J. G. Fossum: IEEE Trans. Electron Devices **52** (2005) 263.

29) S. Dhar, H. Kosina, V. Palankovski, S. E. Ungersboeck, and S. Selberherr: IEEE Trans. Electron Devices **52** (2005) 527.

30) W. Y. Choi, B. G. Park, J. D. Lee, and T. J. K. Liu: IEEE Electron Device Lett. **28** (2007) 743.

31) M. J. Kumar and S. Saurabh, 2008 NSTI Nanotechnology Conference and Trade Show, June 1-5, 2008, Boston, MA, U.S.A.

**Table 1 Device Parameters used in our simulations.**

| | |
|---|---|
| Ge mole fraction in the SiGe buffer layer, x | 0-0.5 |
| Source doping (atoms/cm$^3$) | $1 \times 10^{20}$ |
| Drain doping (atoms/cm$^3$) | $5 \times 10^{18}$ |
| Channel doping (atoms/cm$^3$) | $1 \times 10^{17}$ |
| Channel length, L (nm) | 50 |
| Gate oxide thickness, $t_{ox}$ (nm) | 3 |
| Strained silicon body thickness, $t_{Si}$ (nm) | 10 |
| Gate work function (eV) | 4.5 |
| Drain bias, $V_{DS}$ (V) | 1 |

# <u>List of Figures</u>

Fig. 1: Cross-sectional view of strained double gate tunnel field effect transistor.

Fig. 2: (a) Band diagram of an SDGTFET (x = 0.0) taken along the X-axis at a distance of 2 nm from the oxide-silicon interface for $V_{GS} = 0$ V, $V_{DS} = 1$ V and $V_{GS} = 1$ V, $V_{DS} = 1$ V, (b) Band diagram of an SDGTFET (x = 0.5) taken along the X-axis at a distance of 2 nm from the oxide-silicon interface for $V_{GS} = 0$ V, $V_{DS} = 1$ V and $V_{GS} = 1$ V, $V_{DS} = 1$ V, and (c) Extracted tunnel width on the source side for an SDGTFET at $V_{GS} = 1$ V, $V_{DS} = 1$ V at different Ge mole fractions.

Fig. 3: Transfer characteristics of an unstrained DGTFET and DGFET. ITRS near-term guideline for low standby power technology is also marked in the figure. (Device parameters as in Table 1.)

Fig. 4: Transfer Characteristics of DGTFET and SDGTFET at different strains. ITRS near-term guideline for low standby power technology is also marked in the figure. (Device parameters as in Table 1.)

Fig. 5: (a) Ratio of on-current to off-current of an SDGTFET versus Ge mole fraction for different $V_{GS}$. (Device parameters as in Table 1.) and (b) Band diagram of a DGTFET (SDGTFET, x=0.0) and SDGTFET (x = 0.5) taken along the X-axis at a distance of 2 nm from the oxide-silicon interface for $V_{GS} = 0$ V, $V_{DS} = 1$ V.

Fig. 6: Transfer characteristics of a DGTFET. The only parameter that is changed in the simulation model is the mobility in order to study its effect. (Device parameters as in Table 1.)

Fig. 7: Transfer characteristics of an SDGTFET (Ge mole fraction = 0.5) when operating as both PMOS and NMOS type. For a PMOS type SDGTFET, the $n^+$ (source) doping is $1 \times 10^{20}$

atoms/cm$^3$ and p$^+$ (drain) doping is 5x10$^{18}$ atoms/cm$^3$. Other device parameters are same as in Table 1 for both these devices.

Fig. 8: Threshold voltage of an SDGTFET versus Ge mole fraction for different work function. (Device parameters as in Table 1.)

Fig. 9: (a) Ratio of on-current to off-current versus Ge mole fraction for different gate work function. (Device parameters as in Table 1.) and (b) Combination of Ge mole fraction and gate work function that can be used to get the required threshold voltage while maintaining the ratio of on-current and off-current greater than 1x10$^8$. (Device parameters as in Table 1.)

Fig. 10: Average subthreshold slope of an SDGTFET versus Ge mole fraction. (Device parameters as in Table 1.)

Fig. 11: Threshold voltage versus channel length of DGTFET and SDGTFET for different Ge mole fractions. (Device parameters as in Table 1.). The threshold voltage versus channel length is also shown for a DGFET of similar dimension (silicon body thickness = 10 nm, gate oxide thickness = 3 nm, $\Phi_m$=4.71 eV).

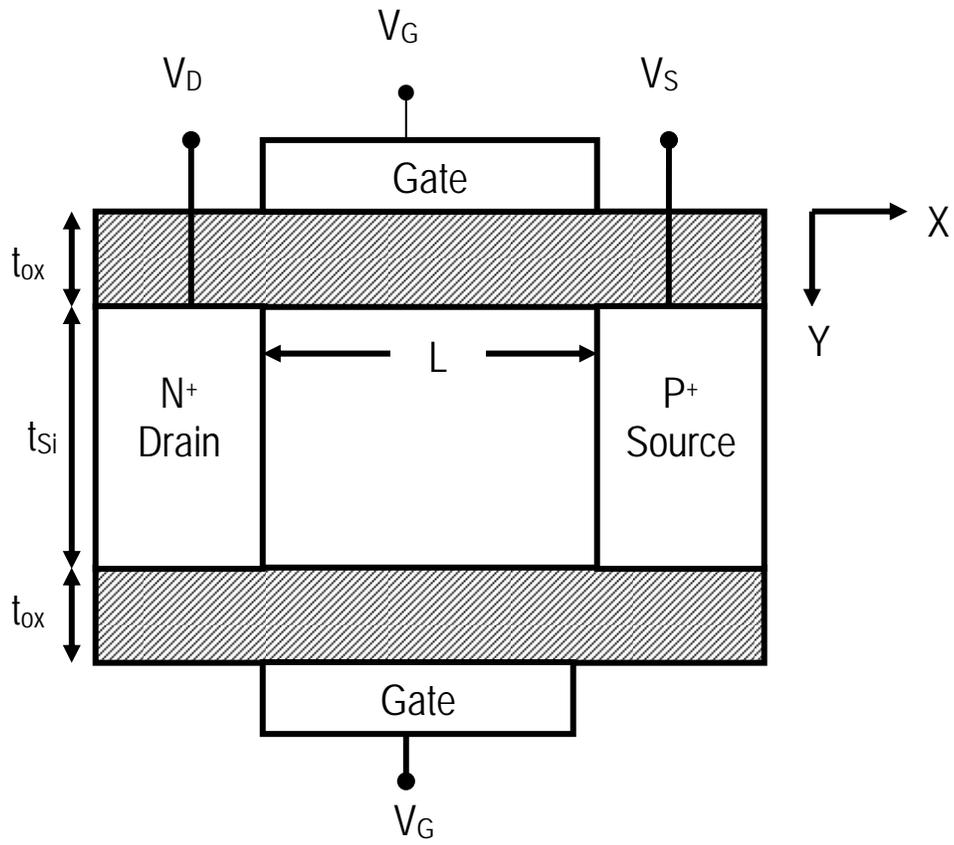

Fig. 1

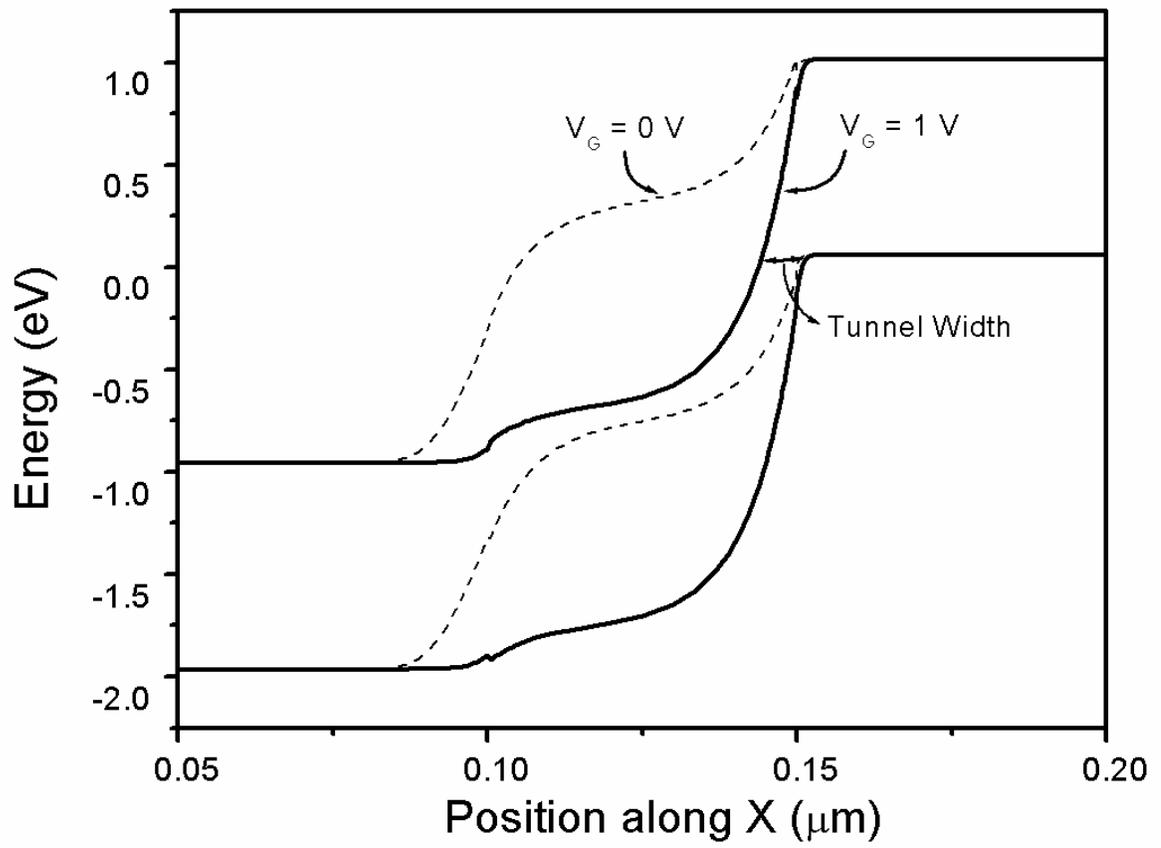

Fig. 2(a)

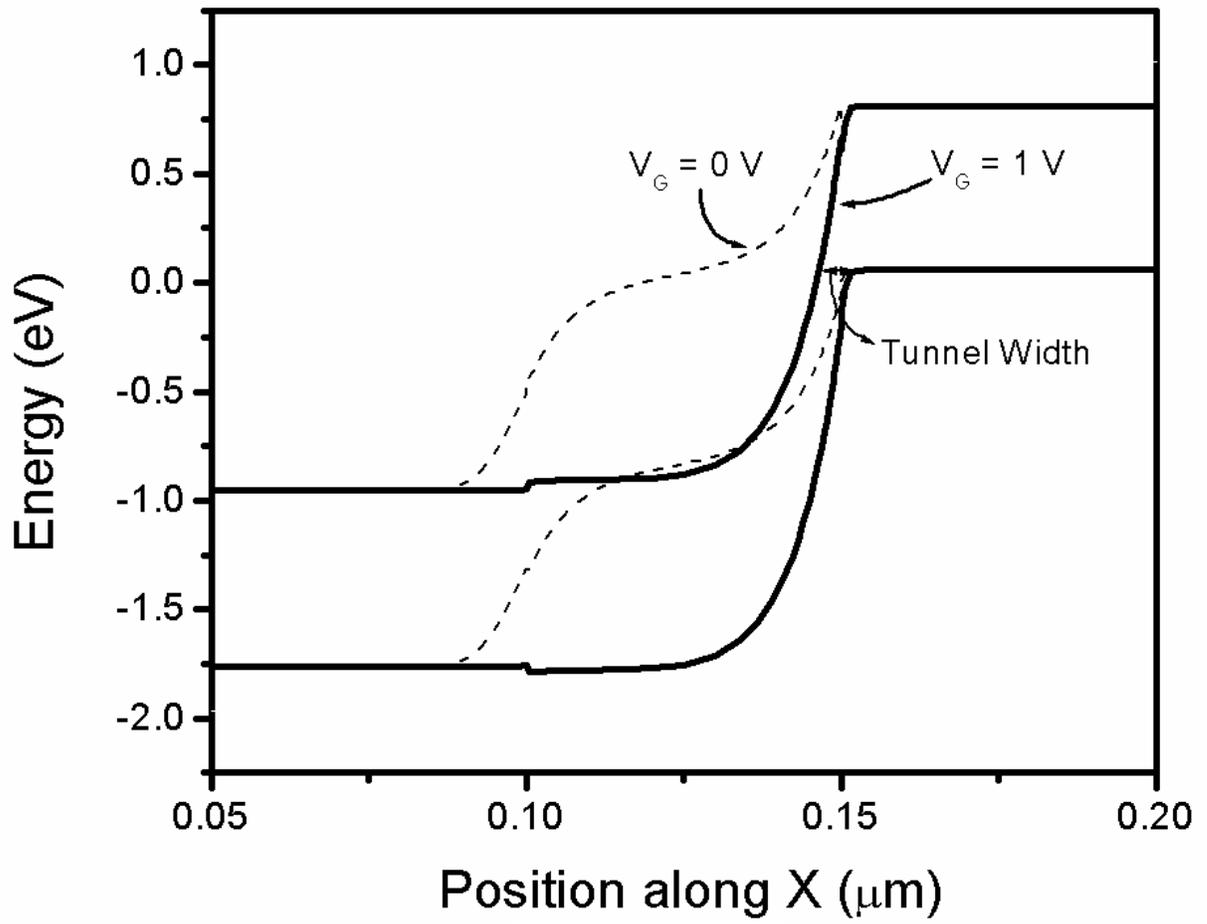

Fig. 2(b)

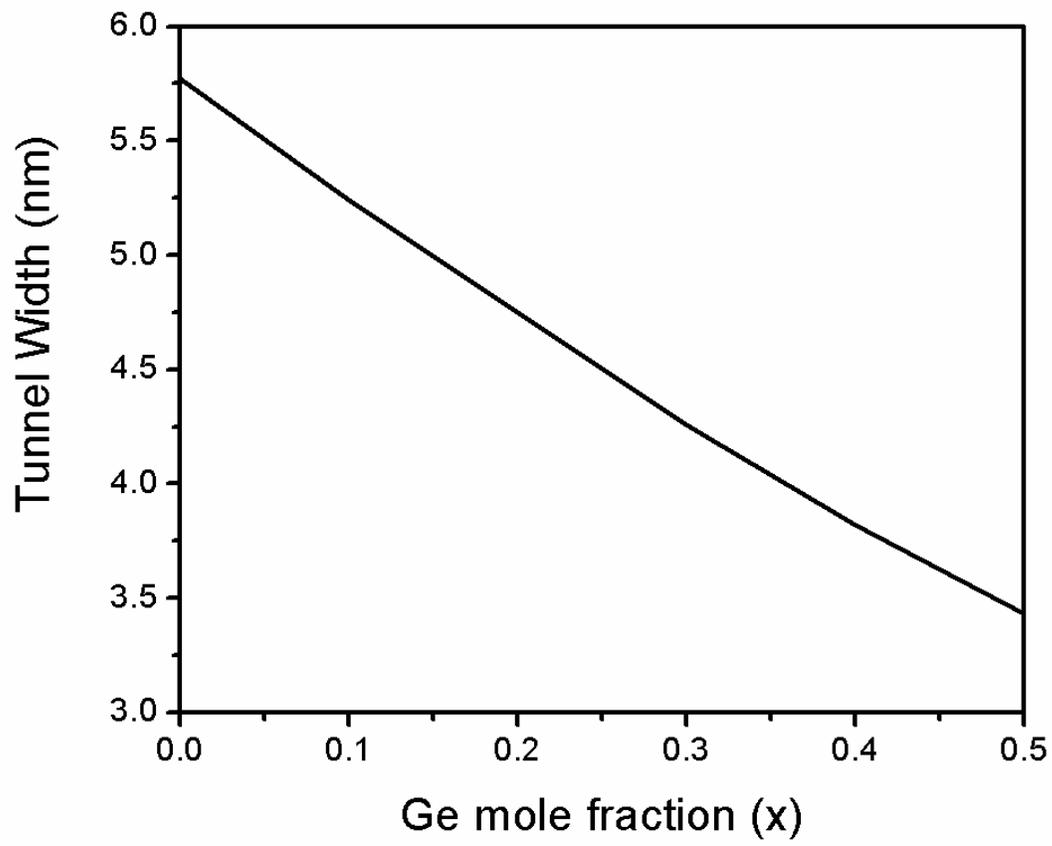

Fig. 2(c)

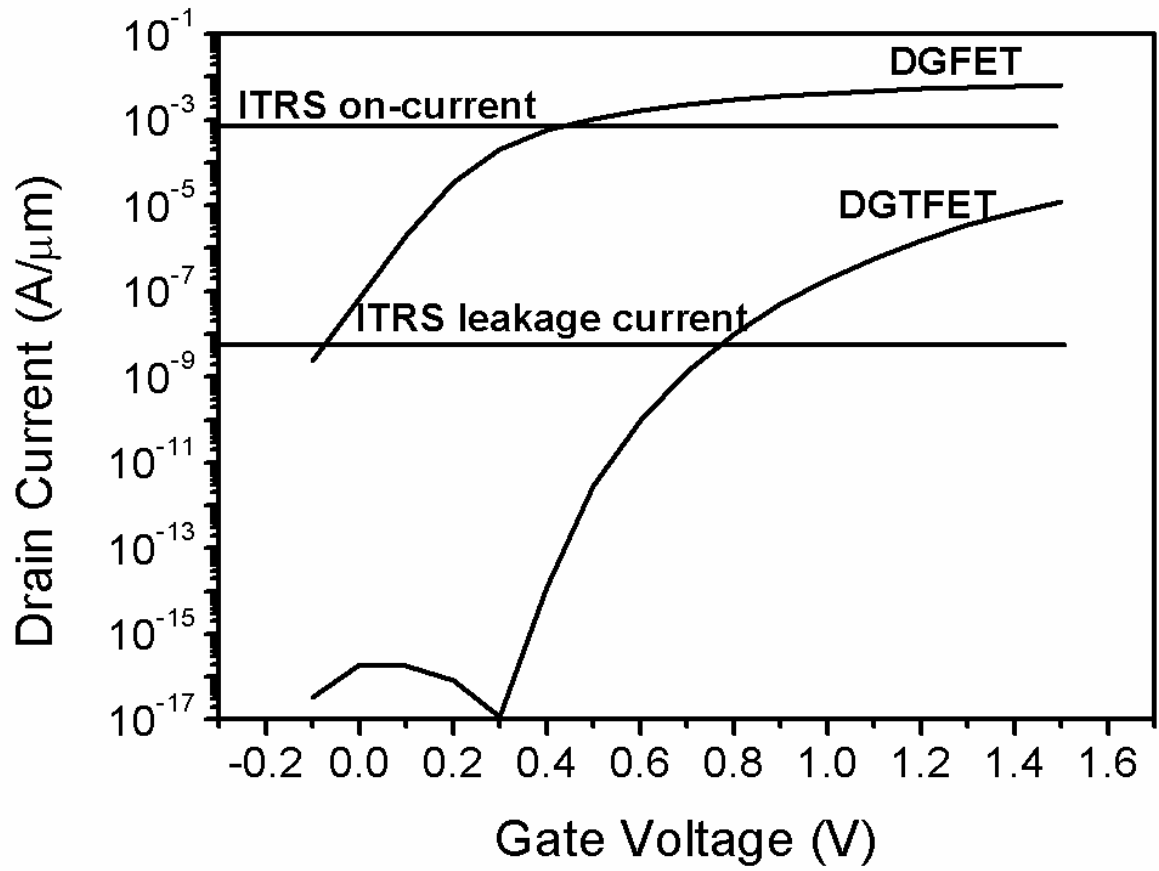

Fig. 3

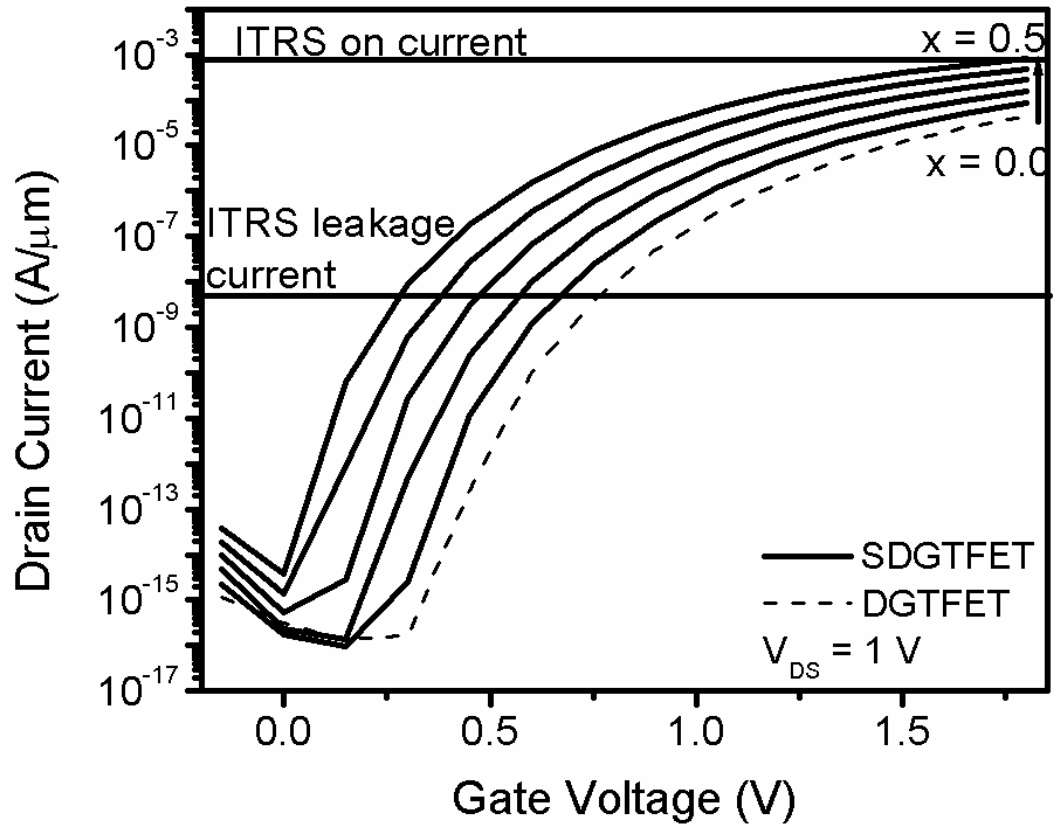

Fig. 4

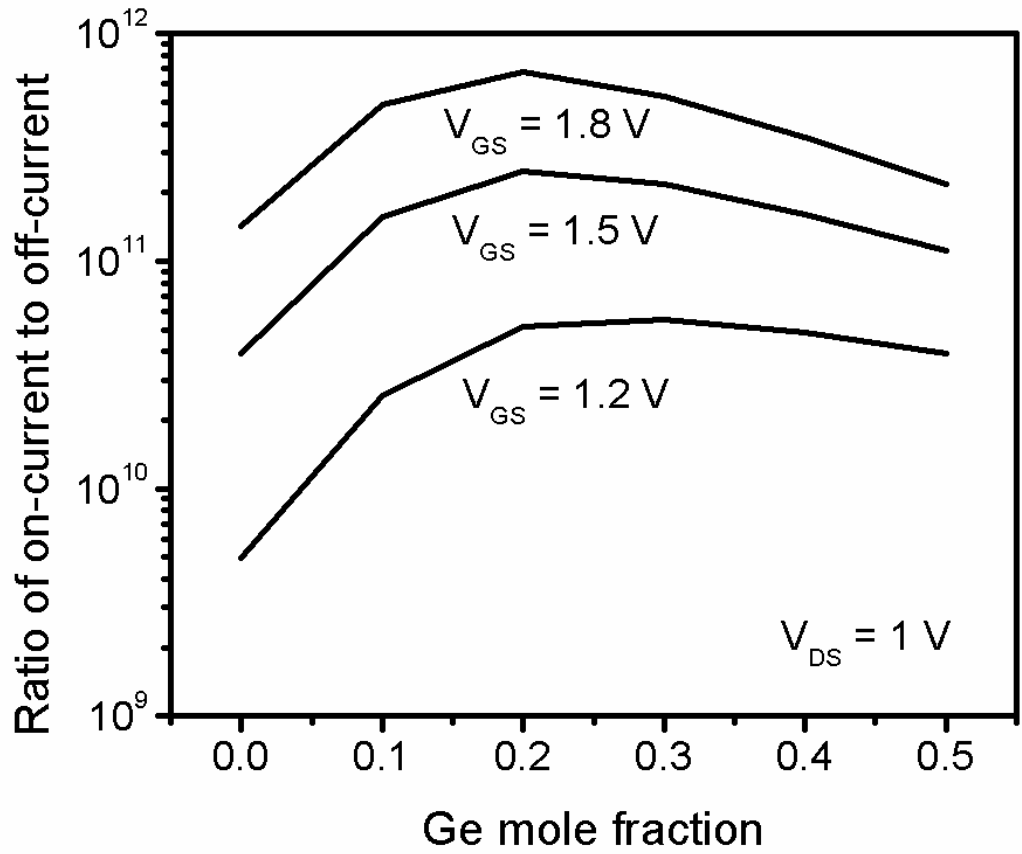

Fig. 5(a)

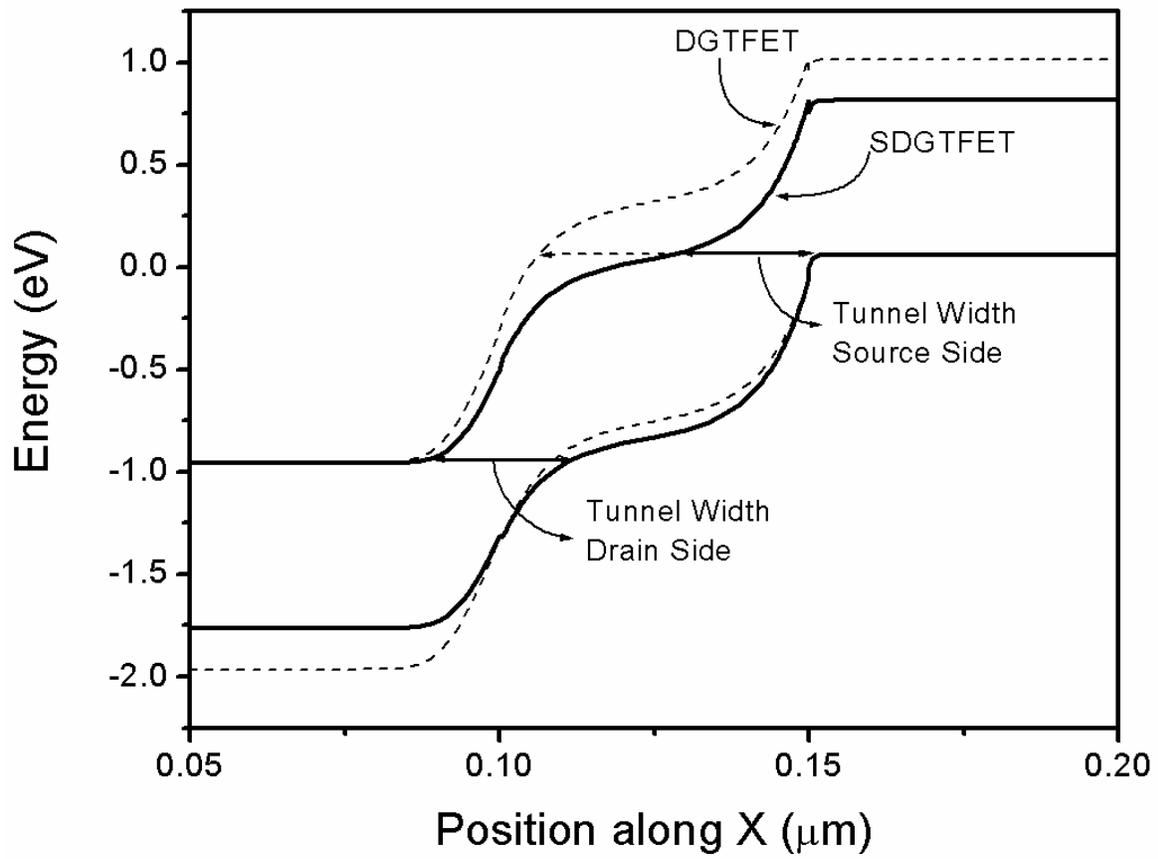

Fig. 5(b)

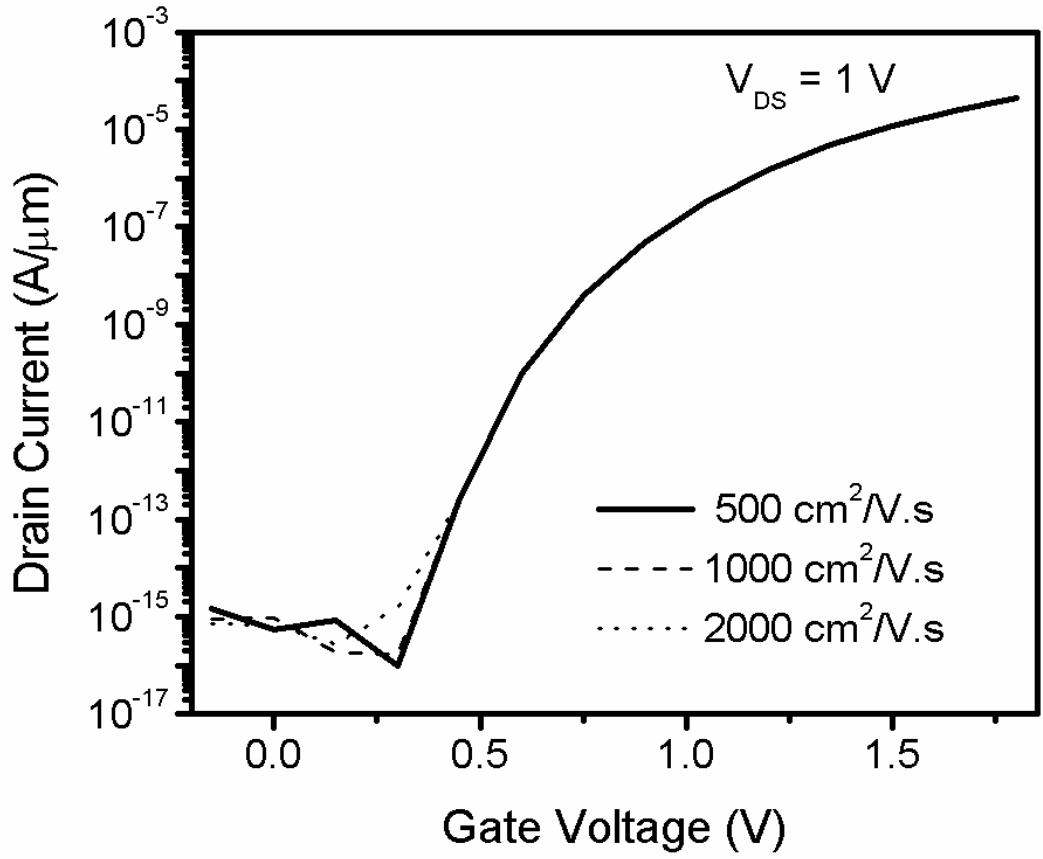

Fig. 6

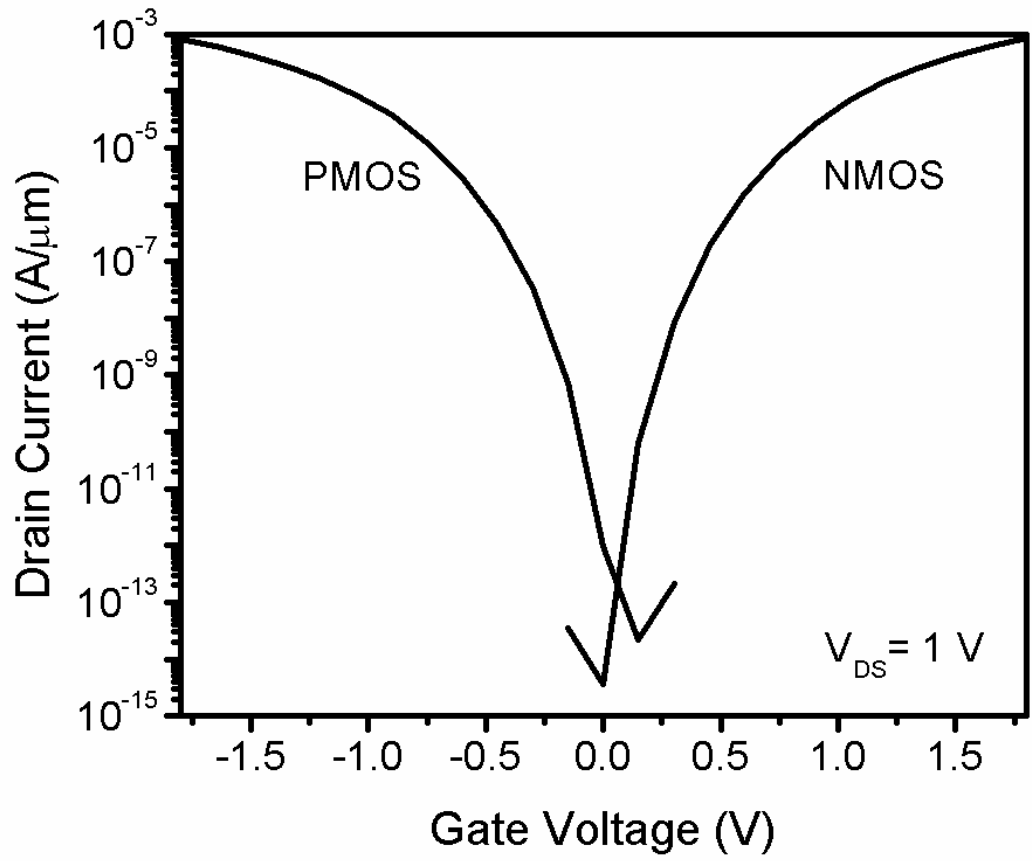

Fig. 7

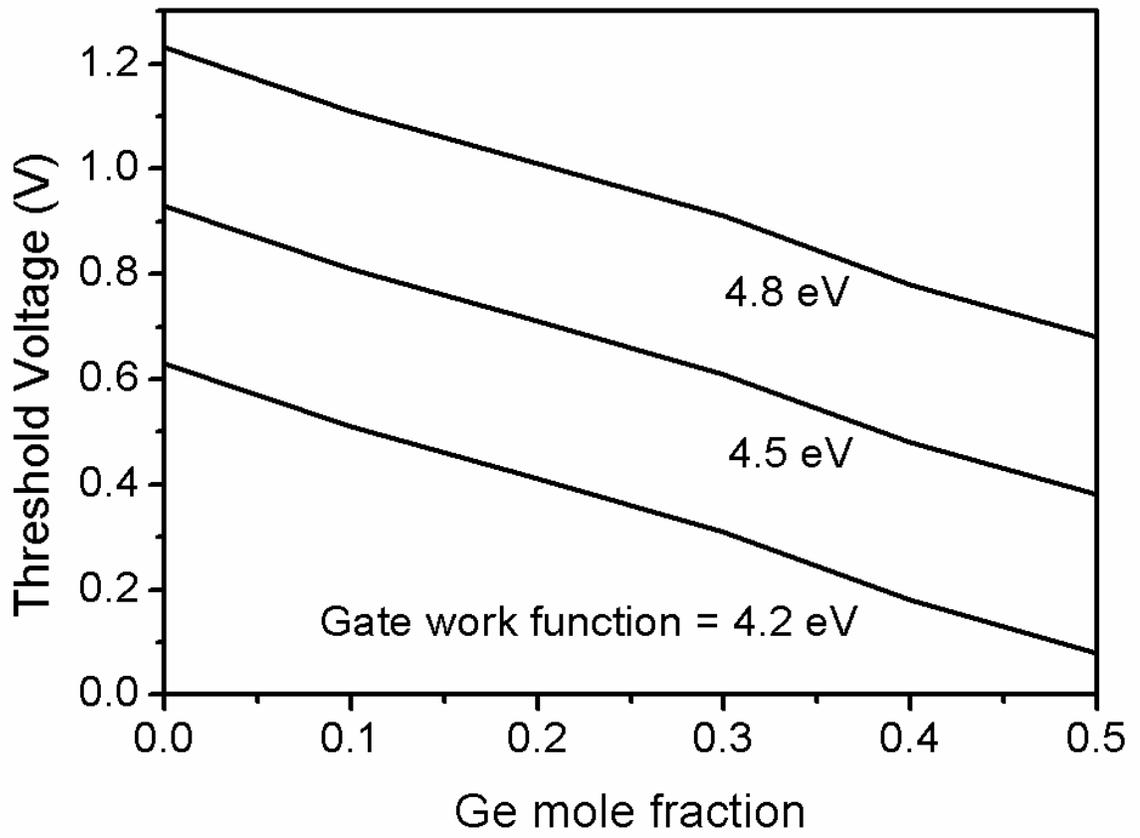

Fig. 8

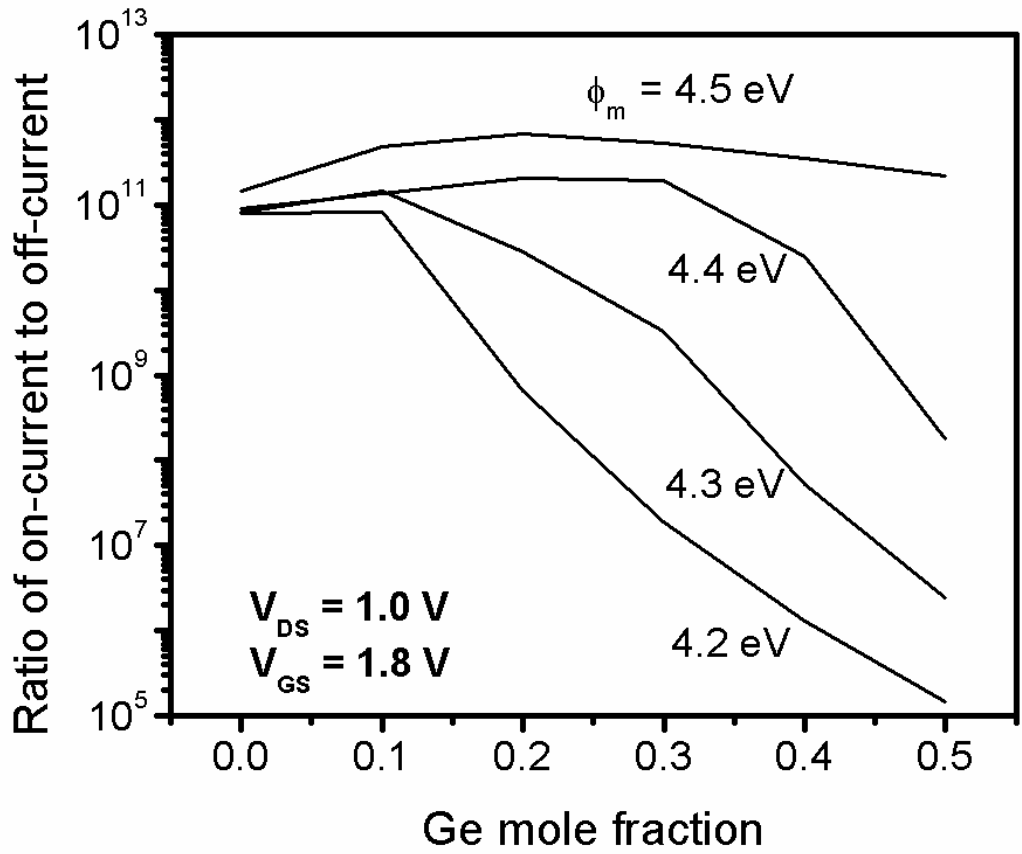

Fig. 9(a)

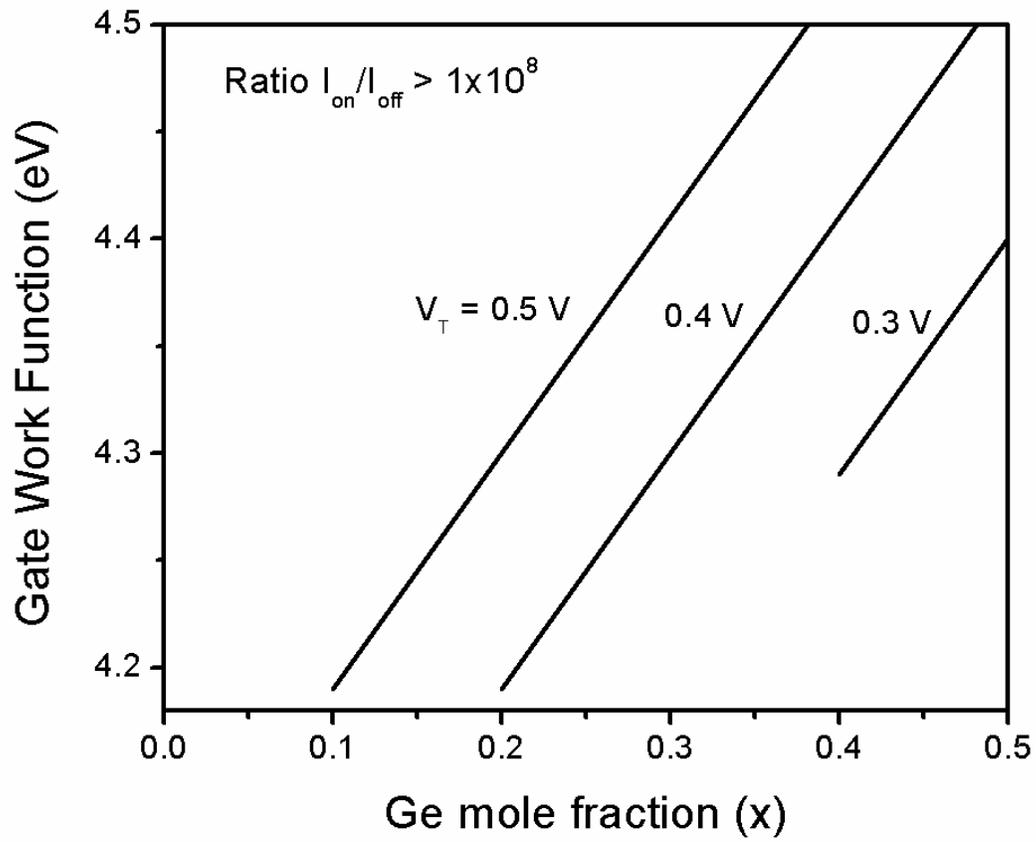

Fig. 9(b)

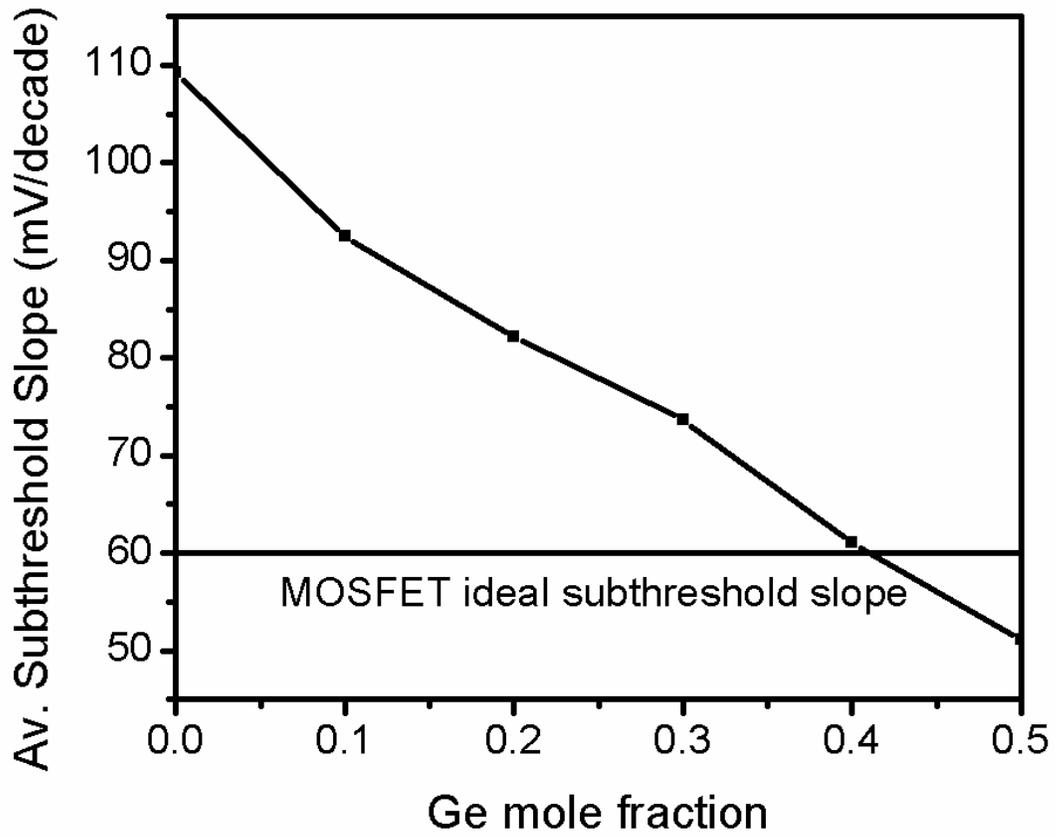

Fig. 10

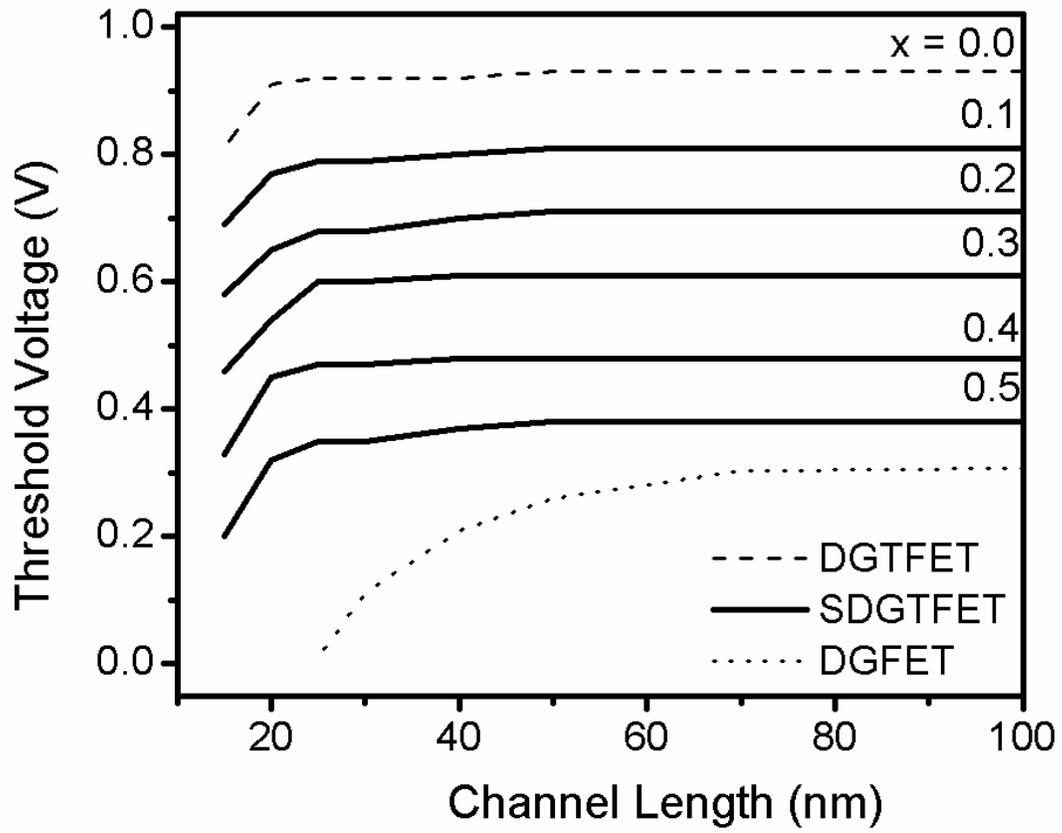

Fig. 11